\documentclass[twocolumn,showpacs,preprintnumbers,amsmath,amssymb]{revtex4}

\usepackage{graphicx}
\usepackage{dcolumn}
\usepackage{bm}

\begin{document}

\title{Information-Theoretic Uncertainty Relation and Random-Phase Entropy}

\author{Kyoung Kon Kim}
\author{Sang Pyo Kim}\email{sangkim@kunsan.ac.kr}
\affiliation{Department of Physics, Kunsan National University,
Kunsan 573-701, Korea}
\author{Sok Kuh Kang}
\affiliation{Korea Ocean Research \& Development Institute, Ansan
426-744, Korea}

\date{\today}
\begin{abstract}
Dunkel and Trigger [Phys. Rev. A {\bf 71}, 052102 (2005)] show that
the Leipnik's joint entropy monotonously increases for the initially
maximally classical Gaussian wave packet for a free particle. After
expressing the joint entropy of the general Gaussian wave packets
for quadratic Hamiltonians as $S (t) = \ln (e/2) + \ln (2 \Delta x
(t) \Delta p (t)/\hbar)$, we show that a class of general Gaussian
wave packets does not warrant the monotonous increase of the joint
entropy. We propose that the random-phase entropy with respect to
the squeeze angle always monotonously increases even for
non-maximally classical states.
\end{abstract}

\maketitle

\section{Introduction}

The Leipnik's joint entropy is an important measure of information
loss of quantum states. Recently Dunkel and Trigger
\cite{dunkel-trigger} have shown that the Leipnik's joint entropy
monotonously increases for a free Gaussian wave packet and a wave
packet in a monochromatic electromagnetic field when it is initially
prepared as a maximally classical state. They have further argued
that the monotonous increase of the entropy may be an intrinsic
property for other types of initial wave packets. However, it was
shown in Ref. \cite{garbaczewski05} that the Leipnik's joint entropy
for a Gaussian wave packet of a harmonic oscillator oscillates
between the maximum and the lower bound, which was also confirmed in
Ref. \cite{OAS08}.

In the present paper, we study the Leipnik's joint entropy for the
general Gaussian wave packets of quadratic Hamiltonians, whose mass
and frequency may depend on time, and which may be driven by an
external force. The exact Gaussian wave packets involve the initial
position and momentum for the centroid, and the initial position and
momentum variances, which are equivalent to a squeeze parameter and
a squeeze angle \cite{kim06}. We first express the Leipnik's joint
entropy in terms of the uncertainty relation and then investigate
the condition for the monotonous increase of the entropy. The joint
entropy of general Gaussian wave packets for a harmonic oscillator
oscillates between the maximum and the lower bound of the joint
entropy, while a class of Gaussian wave packets of a free particle
has the joint entropy that first decreases and then monotonously
increases. However, the random-phase entropy with respect to the
squeeze angle indeed monotonously increases for the general Gaussian
wave packets for quadratic Hamiltonian systems, including a free
particle, a harmonic oscillator and the Caldirola-Kanai Hamiltonian
with an exponentially varying mass \cite{cal}, regardless of an
external driving force.

\section{Joint Entropy for Gaussian Wave Packets}

We consider the quadratic Hamiltonians of the form
\begin{eqnarray}
H (t) = \frac{p^2}{2m(t)} + \frac{m(t) \omega^2(t)}{2} x^2 - f(t) x
. \label{ham}
\end{eqnarray}
As shown in Ref. \cite{kim-page}, the most general quadratic
Hamiltonian may be canonically transformed into the form
(\ref{ham}). A harmonic oscillator has constant $m_0$ and $\omega_0$
while a free particle is the limit of $\omega =0$. A charged
particle in a monochromatic electric field is described by $f(t) =
qE \cos(\omega t)$. The Hamiltonian with time-dependent mass or
frequency describes an open system. To find the exact Gaussian wave
packet for (\ref{ham}), we adopt the invariant method by Lewis and
Riesenfeld \cite{lewis-riesenfeld} (see also Ref. \cite{JKK}). The
Hamiltonian (\ref{ham}) has a pair of linear invariant operators
\cite{MMT,kim99,kim-lee,nieto-truax}, which leads to the most
general Gaussian wave packet \cite{kim06}
\begin{eqnarray}
\Psi (x, t) &=& \Bigl(\frac{1}{ \sqrt{2\pi \hbar} u^*(t)}
\Bigr)^{1/2} e^{- i S_c(t)/ \hbar} e^{i p_c (t) x/\hbar}
\nonumber\\&& \times e^{- i \frac{m\dot{u}^*}{2 \hbar u^*} (x -
x_c(t))^2}. \label{gauss}
\end{eqnarray}
A direct calculation shows that (\ref{gauss}) indeed satisfies the
time-dependent Schr\"{o}dinger equation. Here, $u$ is a complex
solution to
\begin{eqnarray}
\ddot{u} + \frac{\dot{m}}{m} \dot{u} + \omega^2(t) u  = 0, \label{cl
eq}
\end{eqnarray}
satisfying the Wronskian condition
\begin{eqnarray}
m (u \dot{u}^* - \dot{u} u^*) = i. \label{wr}
\end{eqnarray}
And, $x_c$ is the classical position satisfying
\begin{eqnarray}
\ddot{x}_c + \frac{\dot{m}}{m} \dot{x}_c + \omega^2(t) u_c  = f(t),
\label{cl pos}
\end{eqnarray}
and $p_c (t) = m \dot{x}_c (t)$ the corresponding momentum, and $S_c
(t) = \int_0^{t} dt (p_c^2/2m - m \omega^2 x_c^2/2 + f x_c)$ the
classical action. The two variances are given by
\begin{eqnarray}
\Delta (x)(t) &=&\sqrt{\langle x^2 \rangle - \langle x \rangle^2}
= \sqrt{\hbar u^* (t) u(t)}, \nonumber\\
\Delta (p)(t) &=& \sqrt{\langle p^2 \rangle - \langle p \rangle^2} =
m \sqrt{\hbar \dot{u}^* (t) \dot{u} (t)}.
\end{eqnarray}

The Leipnik's joint entropy
\begin{eqnarray}
S (t) &=& - \int dx |\psi (x,t)|^2 \ln |\psi (x, t)|^2 \nonumber\\&&
- \int dp |\tilde{\psi} (p,t)|^2 \ln |\tilde{\psi} (p, t)|^2 - \ln
(2 \pi \hbar) \label{leip}
\end{eqnarray}
of the Gaussian wave packet (\ref{gauss}) is determined by the
uncertainty relation as
\begin{eqnarray}
S (t) = \ln \Bigl( \frac{e}{2} \Bigr) + \ln \Bigl(\frac{2 \Delta x
(t) \Delta p (t))}{\hbar} \Bigr). \label{ent-unc}
\end{eqnarray}
Note that the centroid $(x_c, p_c)$ does not give any contribution
to the joint entropy. So, the question is how to find the variances
for the position and momentum. There is arbitrariness in choosing a
complex solution $u$ to Eq. (\ref{cl eq}) since any linear
superposition of $u$ and $u^*$ is also a solution. For instance, the
Gaussian wave packet for a free particle with the minimum
uncertainty at $t = 0$ is obtained by the solution
\begin{eqnarray}
u_0(t) = \frac{1}{\sqrt{2}} \Bigl(1- i \frac{t}{m_0} \Bigr),
\end{eqnarray}
and for a harmonic oscillator by
\begin{eqnarray}
u_0(t) = \frac{1}{\sqrt{2 m_0 \omega_0}} e^{-i \omega_0 t}.
\end{eqnarray}
Then, the most general solution satisfying Eq. (\ref{wr}) may
involve two parameters $r$ and $\vartheta$ as
\begin{equation}
u (t) = (\cosh r) u_0(t) + (e^{-i \vartheta} \sinh r) u_0^* (t),
 \label{gen cl}
\end{equation}
where $r \geq 0$ and $2 \pi > \vartheta \geq 0$. Here, the $r$ and
$\vartheta$ have the interpretation as the squeeze parameter and
angle \cite{kim99,kim-lee}, which are equivalent to determining two
integration constants, $\Delta x (0)$ and $\Delta p (0)$. In fact,
the general solution (\ref{gen cl}) leads to the Bogoliubov
transformation between the Fock bases constructed from $u(t)$ and
$u_0(t)$:
\begin{eqnarray}
a (t) &=& (\cosh r) a_0 (t) - (e^{i \vartheta} \sinh r)
a_0^{\dagger} (t),
\nonumber\\
a^{\dagger} (t) &=& (\cosh r) a_0^{\dagger} (t) - (e^{-i \vartheta}
\sinh r) a_0 (t).
\end{eqnarray}
Though the free particle allows a shift of time, which corresponds
to choosing a set of $r$ and $\vartheta$, the Hamiltonian
(\ref{ham}) does not preserve the time-translational symmetry, so we
select the wave packets at $t= 0$ by adjusting $r$ and $\vartheta$.

First, we consider the joint entropy for the free particle with
$\omega = 0$. As the external force only governs the centroid of the
Gaussian wave packet, the joint entropy does not depend on the
details of the external force. We thus obtain the Leipnik's joint
entropy
\begin{eqnarray}
&&S (t) = \ln \Bigl( \frac{e}{2} \Bigr) + \frac{1}{2} \ln
\Bigl(\cos^2 \frac{\vartheta}{2} + e^{4r} \sin^2 \frac{\vartheta}{2}
\Bigr) + \nonumber\\&& \frac{1}{2} \ln \Bigl((\cos
\frac{\vartheta}{2} +\frac{t}{m_0} \sin \frac{\vartheta}{2})^2 +
e^{-4r} (\sin \frac{\vartheta}{2} -\frac{t}{m_0} \cos
\frac{\vartheta}{2})^2 \Bigr). \nonumber\\
\label{leip fr en}
\end{eqnarray}
The initial entropy at $t =0$
\begin{eqnarray}
S (0) = \ln \Bigl( \frac{e}{2} \Bigr) + \frac{1}{2} \ln(1+ \sin^2
\vartheta \sinh^2 2r) \label{in en}
\end{eqnarray}
depends not only on the squeeze parameter $r$ but also the squeeze
angle $\vartheta$, which is drawn in Fig. 1.
\begin{figure}
\includegraphics[scale=0.6]{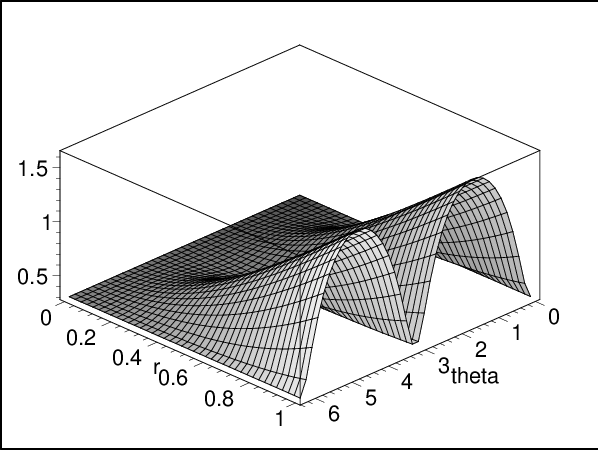}
\caption{The entropy change $S(0) - \ln(e/2)$ for the free particle
as the function of the squeeze parameter $r$ and the squeeze angle
$\vartheta$ is drawn in the ranges $1 \geq r \geq 0$ and $2 \pi >
\vartheta \geq 0$.}
\end{figure}
One interesting point is that the minimum entropy may be provided by
the zero-squeeze angle $(\vartheta = 0)$ even for the non-zero
squeeze parameter $r$. The entropy (\ref{leip fr en}) monotonously
increases against time in the range of squeeze angle $(\pi \geq
\vartheta \geq 0)$ and has the minimum entropy $S(0)$. This is
numerically shown in Fig. 2.
\begin{figure}
\includegraphics[scale=0.6]{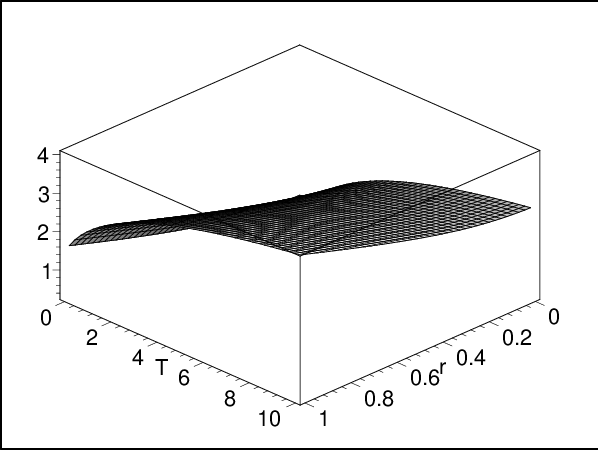}
\caption{The entropy change $S(t) - \ln(e/2)$ for the free particle
with $\vartheta = \pi/2$ as the function of time $T = t/m_0$ and the
squeeze parameter $r$.}
\end{figure}

However, in the other range $(2 \pi > \vartheta > \pi)$ the entropy
has the lower bound
\begin{eqnarray}
S_{\rm min} = \ln \Bigl( \frac{e}{2} \Bigr),
\end{eqnarray}
at $t_* = - (1 - e^{-4r})\sin \vartheta/2m(\sin^2(\vartheta/2) +
e^{-4r} \cos^2 (\vartheta/2))$, which is numerically confirmed in
Fig. 3. Thus, the Gaussian wave packets for (\ref{ham}) do not
necessarily warrant the monotonous increase of the joint entropy.
This counter-example implies that the argument by Dunkel and Triger
is only true for the initially maximally classical states with
$\Delta x (0) \Delta p(0) = \hbar /2$ \cite{dunkel-trigger}. Indeed,
the initially maximally classical state is provided either by
$\vartheta = 0, \pi $ for any $r$, leading to the monotonously
increasing entropy
\begin{eqnarray}
S(t) = \ln \Bigl( \frac{e}{2} \Bigr) + \frac{1}{2} \ln \Bigl(1+
e^{\pm 4r} \frac{t^2}{m^2_0} \Bigr),
\end{eqnarray}
with the upper/lower sign for $\vartheta = \pi/ 0$ or by $r = 0$ for
any $\vartheta$, leading to another monotonously increasing entropy
\begin{eqnarray}
S(t) = \ln \Bigl( \frac{e}{2} \Bigr) + \frac{1}{2} \ln \Bigl(1+
\frac{t^2}{m^2_0} \Bigr).
\end{eqnarray}
\begin{figure}
\includegraphics[scale=0.6]{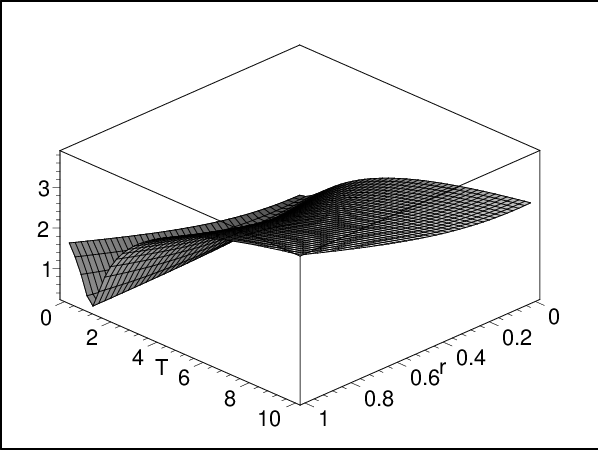}
\caption{The entropy change $S(t) - \ln(e/2)$ for the free particle
with $\vartheta = 3\pi/2$ as the function of time $T = t/m_0$ and
the squeeze parameter $r$. The entropy reaches the lower bound $\ln
(e/2)$ at $t_*$.}
\end{figure}

Second, we consider the joint entropy of the general Gaussian wave
packets for the harmonic oscillator with $m_0$ and $\omega_0$. The
Leipnik joint entropy is given by
\begin{eqnarray}
S (t) = \ln \Bigl(
\frac{e}{2} \Bigr) + \frac{1}{2} \ln \Bigl(1+ \sinh^2 (2r) \sin^2 (2
\omega_0 t - \vartheta) \Bigr). \label{leip os en}
\end{eqnarray}
As shown in Fig. 4, the joint entropy oscillates between the maximum
\begin{eqnarray}
S_{\rm max} = \ln \Bigl( \frac{e}{2} \Bigr) + \frac{1}{2} \ln
\Bigl(1+ \sinh^2 (2r) \Bigr)
\end{eqnarray}
and the lower bound $\ln (e/2)$. This oscillator behavior was first
pointed out in Ref. \cite{garbaczewski05}.
\begin{figure}
\includegraphics[scale=0.6]{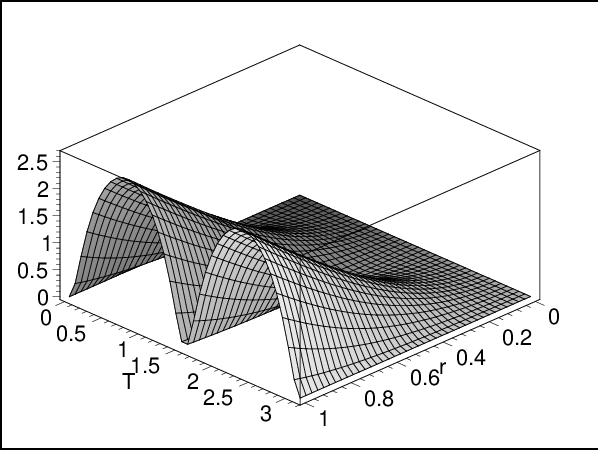}
\caption{The entropy change $S(t) - \ln(e/2)$ for the harmonic
oscillator is drawn as the function of time $T = \omega_0 t$ and the
squeeze parameter $r$ for $\vartheta = 0$.}
\end{figure}

Thus, we may conclude that the Leipnik's joint entropy for the free
particle and the harmonic oscillator does not always monotonously
increase for the whole range of the squeeze parameter and squeeze
angle. Then, a question may be raised whether one may get an
information-theoretic entropy in concord with thermodynamics.

\section{Random-Phase Entropy}

Now we show that the random-phase entropy always leads to a
monotonously increasing entropy for all the Hamiltonians (\ref{ham})
with time-dependent mass or frequency or with the external force.
Regarding the squeeze angle $\vartheta$ as a random variable for the
Gaussian wave packets at $t$, the random-phase entropy is given by
\begin{eqnarray}
\overline{S} (t)  &=& \ln \Bigl( \frac{e}{2} \Bigr) + \ln \Bigl(
\frac{\cosh (2r) +1}{2} \Bigr) \nonumber\\ &&+ \ln \Bigl (2m(t) |u_0
(t) \dot{u}_0 (t)| \Bigr). \label{ran en}
\end{eqnarray}
In deriving Eq. (\ref{ran en}), we used the integral $\int_0^{2 \pi}
\ln (a + b \cos x + c \sin x) = 2 \pi \ln ((a+ \sqrt{a^2 - b^2 -
c^2})/2)$ \cite{PBM}. The random-phase entropy may be written in
terms of the minimal uncertainty as
\begin{eqnarray}
\overline{S} (t)  &=& \ln \Bigl( \frac{e}{2} \Bigr) + \ln \Bigl(
\frac{\cosh (2r) +1}{2} \Bigr) \nonumber\\ &&+ \ln \Bigl ( \frac{2
(\Delta x)_{\Psi_0} (\Delta p)_{\Psi_0}}{\hbar} \Bigr). \label{ran
en2}
\end{eqnarray}
Then, the random-phase entropy is
\begin{eqnarray}
\overline{S} (t)  &=& \ln \Bigl( \frac{e}{2} \Bigr) + \ln \Bigl(
\frac{\cosh (2r) +1}{2} \Bigr) \nonumber\\ &&+ \ln \Bigl ( 1+
\frac{t^2}{m_0^2} \Bigr)
\end{eqnarray}
for the free particle and
\begin{eqnarray}
\overline{S}  = \ln \Bigl( \frac{e}{2} \Bigr) + \ln \Bigl(
\frac{\cosh (2r) +1}{2} \Bigr)
\end{eqnarray}
for the harmonic oscillator.

As an open system, we consider the time-dependent mass $m(t) = m_0
e^{\gamma t}$, which is well-known as the Caldirola-Kanai
Hamiltonian \cite{cal}. The solution to Eq. (\ref{cl eq}) that
satisfies (\ref{wr}) and gives the minimum uncertainty is
\cite{kim03}
\begin{eqnarray}
u_0 (t) = \frac{e^{-\gamma t/2}}{\sqrt{2 m_0 \omega}} e^{-i \omega
t}, \quad \omega = \sqrt{\omega_0^2 - \Bigl(\frac{\gamma}{2}
\Bigr)^2}.
\end{eqnarray}
Then, the random-phase entropy is
\begin{eqnarray}
\overline{S}  = \ln \Bigl( \frac{e}{2} \Bigr) + \ln \Bigl(
\frac{\cosh (2r) +1}{2} \Bigr) + \frac{1}{2} \ln \Bigl(1 +
\frac{\gamma^2}{4 \omega^2} \Bigr).
\end{eqnarray}
Even for $r=0$, the damping effect contributes to the lower bound of
entropy by the amount of the last term, which is related with the
generalized minimum uncertainty \cite{kim03}.

For time-dependent Hamiltonians (\ref{ham}), we may find a useful
lower and upper bound. Using the identity, $\ln (2 m(t) |u_0 (t)
\dot{u}_0 (t)|) =  \ln  ( 1+ (m(t) d|u_0 (t)|^2/dt)^2)/2$, we find
the lower bound for the random-phase entropy
\begin{eqnarray}
\overline{S} (t)  \geq \ln \Bigl( \frac{e}{2} \Bigr) + \ln \Bigl(
\cosh (2r) +1 \Bigr). \label{ran en lo bo}
\end{eqnarray}
Similarly, from $ \langle H(t) \rangle_{\Psi_0} \geq \hbar \omega
(t) m (t) |u_0(t) \dot{u}_0| $, with respect to the Gaussian wave
packet $\Psi_0$ obtained using $u_0(t)$, we find the upper bound
\begin{eqnarray}
 \ln \Bigl( \frac{e}{2} \Bigr) + \ln \Bigl(
\cosh (2r) +1 \Bigr) + \frac{1}{2} \ln \Bigl (\frac{\langle H(t)
\rangle_{\Psi_0}}{\hbar \omega (t)/2} \Bigr) \geq \overline{S} (t).
\label{ran en up bo}
\end{eqnarray}

\section{Conclusion}

We have investigated the Leipnik's joint entropy of the general
Gaussian wave packets for quadratic Hamiltonians. The Gaussian wave
packets (\ref{gauss}) have four integration constants: two constants
for the classical position and momentum, and two variances for the
position and momentum. The first two constants are equivalent to a
complex parameter for the coherent state and the second two
constants to the squeeze parameter and the squeeze angle. The joint
entropy (\ref{ent-unc}) is expressed entirely in terms of the
uncertainty relation, which is independent of the centroid but
depends on the position and momentum variances. We computed the
joint entropy for a free particle, a harmonic oscillator and the
Caldirola-Kanai Hamiltonian with an exponentially varying mass. The
joint entropy depends only on the squeeze parameter and the squeeze
angle as shown in Eqs. (\ref{leip fr en}) for the free particle and
(\ref{leip os en}) for the harmonic oscillator.

The general Gaussian wave packets have the initial entropy (\ref{in
en}) that depends on the squeeze parameter and the squeeze angle. It
is shown that the Leipnik's joint entropy satisfies the lower bound
$S(t) \geq \ln (e/2)$ and monotonously increases for the range of
squeeze angle $\pi \geq \vartheta \geq 0$, while for the remaining
range $2\pi > \vartheta > \pi$ it first decreases until the lower
bound, $\ln (e/2)$, is reached and then it monotonously increases.
This counter-example implies that the joint entropy monotonously
increases only for the initially maximally classical states that
have the minimum uncertainty. Similarly, the joint entropy of the
squeezed vacuum states of the initially maximally classical state
for the harmonic oscillator oscillates between the maximum and the
lower bound, $\ln (e/2)$.

In this paper, we have proposed that the random-phase entropy with
respect to the squeeze angle leads to monotonously increasing
entropy for the general Gaussian wave packets, which is shown
explicitly for a free particle, a harmonic oscillator and the
Caldirola-Kanai Hamiltonian. For a quadratic Hamiltonian with
time-dependent mass or frequency, the joint entropy satisfies the
lower bound (\ref{ran en lo bo}) and the upper bound (\ref{ran en up
bo}). It remains an open question whether the random-phase entropy
monotonously increases for all time-dependent quadratic
Hamiltonians.

\acknowledgements

SPK appreciates APCTP for the warm hospitality, where this paper was
finished. This work has been supported by KORDI research program
(PG47100) through KRCF(Korea Research Council of Fundamental Science
and Technology) of Ministry of Education, Science and Technology.


\begin{references}

\bibitem{dunkel-trigger} J.~Dunkel and S.~A.~Trigger, Phys. Rev. A {\bf
71}, 052102 (2005).

\bibitem{garbaczewski05} P.~Garbaczewski, Phys. Rev. A {\bf 72},
056101 (2005); V.~Majern\'{i}k and T.~Opatrn\'{y}, J. Phys. A: Math.
Gen. {\bf 29}, 2187 (1996).

\bibitem{OAS08} \"{O}.~\"{O}zcan, E.~Akt\"{u}urk, and R.~Sever, Int. J.
Theor. Phys. {\bf 47}, 3207 (2008); E.~Akt\"{u}urk,
\"{O}.~\"{O}zcan, and R.~Sever, Int. J. Mod. Phys. B {\bf 23}, 2449
(2009).

\bibitem{kim06} S.~P.~Kim, J. Korean Phys. Soc. {\bf 49}, 464 (2006)
[arXiv:quant-ph/0511073].

\bibitem{cal} P.~Caldirola, Nuovo Cimento {\bf 18}, 393
(1941); {\bf 77B}, 241 (1983); E.~Kanai, Prog. Theor. Phys. {\bf 3},
440 (1948).

\bibitem{kim-page} S.~P.~Kim and D.~N.~Page, Phys. Rev. A {\bf 64}, 012104 (2001).

\bibitem{lewis-riesenfeld} H.~R.~Lewis, Jr. and W.~B.~Riesenfeld, J. Math.
Phys. {\bf 10}, 1458 (1969).

\bibitem{JKK} J-Y.~Ji, J.~K.~Kim, and S.~P.~Kim, Phys. Rev. A {\bf 51},
4268 (1995); J.-Y.~Ji, J.~K.~Kim, S.~P.~ Kim, and K-S.~Soh, Phys.
Rev. A {\bf 52}, 3352 (1995).

\bibitem{MMT} I.~A.~Malkin, V.~I.~Man'ko, and D.~A.~Trifnov, Phys. Rev. D {\bf
2}, 1371 (1970).

\bibitem{kim99} J.~K.~Kim and S.~P.~Kim, J. Phys. A {\bf 32}, 2711 (1999).

\bibitem{kim-lee} S.~P.~Kim and C.~H.~Lee, Phys. Rev. D {\bf 62}, 125020
(2000).

\bibitem{nieto-truax} M.~M.~Nieto and D.~R.~Truax, New J. Phys. {\bf 2}, 18 (2000).

\bibitem{PBM} I.~S.~Gradshteyn and I.~M.~Ryzhik, {\it Table of
Integrals, Series, and Products} (Academic Press, San Diego, 1994),
p 560, formula 4.225.

\bibitem{kim03} S.~P.~Kim,  J. Phys. A: Math. Gen. {\bf 36},  12089
(2003).

\end{references}
\end{document}